\documentstyle[prl,aps,psfig,epsf,twocolumn]{revtex}
\begin{document}
\draft
\twocolumn[\hsize\textwidth\columnwidth\hsize\csname@twocolumnfalse%
\endcsname

\preprint{}

\title{Bilayer Coherent and Quantum Hall
Phases: Duality and Quantum Disorder}
\author{Eugene Demler$^1$, Chetan Nayak$^2$,
Sankar Das Sarma$^3$}
\address{$^1$Physics Department, Harvard University,
Cambridge, MA 02138\\
$^2$Physics Department, University of California Los Angeles,
Los Angeles, CA 90095\\
$^3$Department of Physics, University of Maryland,
College Park, MD 20742}

\date{\today}
\maketitle
\vskip 0.5 cm
\begin{abstract}
We consider a fully spin-polarized quantum Hall system with
no interlayer tunneling at total filling factor $\nu=1/k$
(where $k$ is an odd integer) using the 
Chern-Simons-Ginzburg-Landau theory. Exploiting particle-vortex
duality and the concept of quantum disordering, we
find a large number of possible compressible and incompressible
ground states, some of which
may have relevance to recent experiments of  
Spielman {\it et.al.} \cite{Spielman00}.
We find interlayer coherent compressible states without
Hall quantization and interlayer-incoherent incompressible
with Hall quantization in addition to the usual $(k,k,k)$
Halperin states, which are both interlayer-coherent
and incompressible.
\end{abstract}

\vspace{1 cm}

\vskip -0.5 truein
\vskip 0.2 cm
]
\narrowtext

Bilayer quantum Hall systems have been a subject of
great experimental and theoretical interest
\cite{Eisenstein97,Fertig89,Wen92,Ezawa92,He93,Moon95,Murphy95,Bonesteel96,DasSarma,Stern00} 
for more than a decade. In particular,
spin-polarized bilayer quantum Hall systems with
little or no inter-layer tunneling (${\Delta_{SAS}}\approx 0$)
at the {\it total} Landau-level filling factor
$\nu=1$ have been studied intensively because the
layer index serves as a pseudospin index
with $U(1)$ symmetry, leading to the possibility
of many interesting quantum phases
and quantum phase transitions associated with the
pseudospin $U(1)$ symmetry. Among the many exotic
possibilities discussed in the literature in this context
are long-range pseudospin order (the so-called spontaneous
interlayer phase coherence) associated with
the spontaneous breaking of the pseudospin $U(1)$ symmetry,
a linearly dispersing collective Goldstone mode,
bilayer pseudospin superfluidity, the associated Josephson
effect, topological defects (merons, skyrmions, and
vortices), and a Kosterlitz-Thouless
phase transition at finite-temperature. The subject is
of considerable current interest as a result of the
recent appearance \cite{Spielman00} of an experimental
paper reporting the possible observation in a bilayer
tunneling measurement of the predicted
Goldstone mode at $\nu=1$ associated with 
pseudospin $U(1)$ symmetry-breaking. There have
been several recent preprints \cite{Schliemann00}
providing possible theoretical explanations of the data
presented in \cite{Spielman00}.

In this letter, we consider a somewhat more general
situation (still with ${\Delta_{SAS}}=0$, although
a generalization to ${\Delta_{SAS}}\neq 0$
is straightforward) with total filling factor
$\nu=1/k$ (with $k$ an odd integer and
$k=1$ being the special situation
primarily discussed in the literature)
from a new theoretical perspective hitherto not considered
in the literature.
We extend the Chern-Simons-Landau-Ginzburg (CSLG) theory for
the Halperin $(k,k,k)$ state \cite{Moon95,Lee90}
to describe the posibility of quantum disordering of the charge
and pseudospin degrees of freedom. Our quantum disordering
procedure relies on the $U(1)$ particle-vortex duality
in $2+1$ dimensions and leads to a large class of incompressible and 
compressible
bilayer states at $\nu=1/k$; some are interlayer
coherent, others are not. One of our important
conclusions 
is that the existence of interlayer coherence is neither
necessary nor sufficient for the existence of an incompressible
bilayer $\nu=1$ (or $1/k$) quantum Hall state (even for
${\Delta_{SAS}}=0$). It is possible to have Hall quantization at
$\nu=1/k$ without having interlayer coherence and vice-versa.
We believe that this important  result of ours 
may have implications for the experimental observations in
ref. \cite{Spielman00}, where the putative Goldstone
mode associated with interlayer coherence
has presumably been observed in a state which is either
compressible or is very weakly incompressible.
We therefore raise the interesting (but, by no means,
definitive) possibility that the bilayer system in
ref. \cite{Spielman00} is {\it not} the Halperin $(1,1,1)$
ground state, as has been universally assumed, in which
interlayer coherence and incompressibility occur
together, but is one of the new quantum disordered
interlayer coherent and compressible states (e.g. the Hall insulator
\cite{Kivelson92} state we find below) predicted in this
paper. The experimental data presented in
ref. \cite{Spielman00} provide some circumstantial evidence
in support of such a tentative claim: (1) the $\nu=1$
zero-bias tunneling conductance peak reported in
\cite{Spielman00} is much broader than the weak ($> 1k\Omega$)
and shallow $\rho_{xx}$ minimum seen in the data
($\rho_{xy}$ quantization has not been observed)
even at the very low temperature of $40 mK$;
(2) the zero-bias tunneling peak width exists 
\cite{Spielman00} over $\Delta \nu\approx 0.7$
and, in fact, contains even the $\rho_{xx}$
maximum within it, raising some doubts about the standard
interpretation of the state as the $(1,1,1)$ state,
which is manifestly an incompressible state;
(3) $\rho_{xx}$ for $\nu<1$ in ref. \cite{Spielman00}
shows striking insulating behavior as it increases
extremely sharply with decreasing $\nu$, indicating
that it may be a Hall insulator for these $\nu$.
We point out that all the quantum disordered
compressible states found by us break translational
invariance. In the clean limit, they are Wigner crystal/charge
density wave/striped phase-type states in each layer.
In the limit that disorder plays a large role
(a more likely scenario for ref. \cite{Spielman00}
where the carrier density is very low and consequently
disorder effects may be important), they are better-described
as disorder-induced localized phases. We believe that some of the bilayer
striped (compressible) interlayer coherent phases recently discussed
in the literature \cite{Brey00} within a microscopic
Hartree-Fock calculation may actually belong to the class
of translational invariance-broken compressible coherent
quantum disordered phases we find below.

Our construction has precedents in the high-$T_c$
superconductivity literature \cite{Balents99},
where it has been used to
describe spin-charge separation. Here, we implement
{\it pseudospin}-charge separation to disorder
the pseudospin and charge currents separately.
In the context of superconductivity (or superfluidity),
it is clear that magnetism and superconductivity (or superfluidity)
are distinct effects.
The triplet order parameter of the $A$-phase of $^3$He
breaks both spin and particle number symmetries,
but these symmetries may be restored separately;
restoring the latter alone would lead to a
magnetic state of solid Helium. Our construction can
be understood in a similar vein, with
the interlayer coherent $(1,1,1)$ quantum Hall state
replacing triplet superconductivity. Our
work is also related to the hierarchy construction
of Lee and Kane \cite{Lee90}, but these authors
only considered the condensation of charged vortices
and skyrmions, which does not lead to our cornucopia
of states with the same total $\sigma_{xy}$.

We begin our discussion of bilayer quantum Hall systems at 
$\nu=1/k$ with a CSLG theory 
\cite{Moon95,Lee90,Zhang89} 
that describes
incompressible interlayer coherent states with the
Halperin $(k,k,k)$
wavefunction. We neglect real spin, assuming
complete spin polarization, but allow unequal layer
populations, i.e. the so-called unbalanced case,
${\rho_1}\neq{\rho_2}$, where ${\rho_i}$ is the electron
density in layer $i$.
The imaginary-time action for this theory is
($\hbar = c = e =1$):
\begin{eqnarray}
\label{L0}
{\cal L} &=& \sum_{i=1,2} \biggl\{ \Psi_i^{\dagger} 
\left( \partial_0 - i a_0 - i A_0^{(i)} \right) 
\Psi_i 
\cr & & 
+ \,\frac{1}{2 m} \left| \left[\vec{\partial}
- i\vec{a} - i\vec{A}^{(i)} \right] \Psi_i \right|^2
 + \,\frac{U}{2} ( \Psi^{\dagger}_i \Psi_i - \rho_i )^2 \biggr\}
\cr
& & + \,V ( \Psi^{\dagger}_1 \Psi_1 - \rho_1 )
( \Psi^{\dagger}_2 \Psi_2 - \rho_2 ) 
\cr & & - \,\frac{i}{4 \pi k} \epsilon_{\mu\nu\lambda}
a_{\mu} \partial_{\nu} a_{\lambda}
\end{eqnarray}
where the $\Psi_i$'s are the bosonic fields which
describe electrons in the two layers and
the final term in (\ref{L0}) is the Chern-Simons
term, ${\cal L}_{CS}(a)$, which enforces
the fermionic statistics of the electrons.
The gauge fields $A_{\mu}^{(1,2)}$ 
couple to electrons in layers 1 and 2 respectively.
By choosing $\rho_1$ and $\rho_2$ to be arbitrary we
allow for charge imbalance between 
the layers. For simplicity, we replace the long-range 
Coulomb interaction by short-range interactions.
Intra-layer interactions are not equal to
inter-layer interactions, $U\neq V$, so the
model has only $U(1)$ pseudospin symmetry,
rather than the full $SU(2)$ which would
be obtained if $U=V$.

We take $\Psi_i = \sqrt{ \rho_i} e^{ i \theta_i}$
and introduce $\theta =  ({\theta_1} + {\theta_2})/2$,
$\phi = ({\theta_1} - {\theta_2})/2$.
Pseudospin $U(1)$ is the symmetry
$\phi\rightarrow \phi +\, {\rm const}$.
When $\phi$ spontaneously chooses a direction,
this symmetry is broken
and the oscillations of $\phi$ describe the
Goldstone mode associated with this broken symmetry.
Similarly, we define
$A_{\mu}^C  = (A^{(1)}_{\mu} + A^{(2)}_{\mu})/2$
and $A_{\mu}^I  = (A^{(1)}_{\mu} - A^{(2)}_{\mu})/2$.
$A^C$ couples to the total charge of the system
while $A^{I}$ couples to the charge difference
between the two layers, i.e. the pseudospin.
Vortices in $\theta_1$ or $\theta_2$ are called merons
\cite{Moon95}. They come in four varieties since they
can be either vortices or antivortices in
$\theta_1$ or $\theta_2$, and they have
charge $\pm {\rho_{1,2}}/[({\rho_1}+{\rho_2})k]$.
Vortices in $\theta_1$ and $\theta_2$
can be combined to form a vortex in
$\theta$. In pseudospin language,
they are skyrmions \cite{Moon95}, carrying charge $\pm 1/k$.
A vortex in $\theta_1$ can
be combined with an anti-vortex
in $\theta_2$ to form a vortex in $\phi$ of
charge $({\rho_1}-{\rho_2})/[({\rho_1}+{\rho_2})k]$.

If we integrate out the amplitude fluctuations
in (\ref{L0}) and set $( \rho_1 + \rho_2 )/ m = 1$,
$K_3 = ( \rho_1 - \rho_2)/ ( \rho_1 + \rho_2)$,
we obtain
\begin{eqnarray}
{\cal L} &=& \frac{1}{2{v_c^2}} \left( \partial_0 \theta -
a_0 -A_0^C \right)^2
 + \frac{1}{2{v_s^2}} \left( \partial_0 \phi  -A_0^I \right)^2
\cr & &
+ \,\frac{1}{2}
\, \left( \vec{\partial} \theta - \vec{a} -\vec{A}^C \right)^2 
+ \,\frac{1}{2}\, \left( \vec{\partial} \phi -\vec{A}^I \right)^2
\cr & &
 + \,K_3 \,\,
\left( \vec{\partial} \theta - \vec{a} -\vec{A}^C \right) 
\cdot\left( \vec{\partial} \phi -\vec{A}^I \right)
\cr & & -\,{\cal L}_{CS}(a)
\label{L1}
\end{eqnarray}
where ${v_{c,s}}$ are the velocities of the
charge and pseudospin collective modes.

Using the standard $U(1)$ particle-vortex 
duality \cite{dual} for 
both $\theta$ and $\phi$ and their vortex
excitations, we rewrite (\ref{L1}) as
\begin{eqnarray}
\tilde{{\cal L}}_D &=& 
\frac{ \rho_C}{2} \left| \left( \partial_{\mu} -
i b_{\mu}^C \right) \Phi_C \right|^2
+\frac{ \rho_I}{2} \left| \left( \partial_{\mu} -
i b_{\mu}^I \right) \Phi_I \right|^2
\cr
& & +\,\frac{\kappa_0}{4} ( F^C_{\alpha\beta} )^2
 - \pi k i \epsilon_{\mu\nu\lambda}  b_{\mu}^C 
\partial_{\nu} b^C_{\lambda}
- i b_{\mu}^C \epsilon_{\mu\nu\lambda} \partial_{\nu} 
 A^C_{\lambda}
\cr
& & +\,\frac{\kappa_0}{4} ( F^I_{\alpha\beta} )^2
- i b_{\mu}^I \epsilon_{\mu\nu\lambda} \partial_{\nu} 
A^I_{\lambda} - \kappa_3 F^C_{0\alpha} F^I_{0\alpha}
\label{Ltd}
\end{eqnarray}
where $\kappa_0 = 1/(1 -K_3^2 )$,
$\kappa_3 = K_3/(1 -K_3^2 )$, and $F^{C,I}_{\alpha\beta}$
is the field strength associated with $b^{C,I}_\mu$.
(To avoid clutter, we set ${v_{c,s}}=1$; they may be
restored by dividing all temporal derivatives by
the appropriate velocity.) The fields in (\ref{Ltd})
are related to those of (\ref{L1}) according to
\begin{eqnarray}
\epsilon_{\mu\nu\lambda} \partial_{\nu}
{b_{\lambda}^C} \equiv {\partial_\mu}\theta - {a_\mu} - {A_\mu^C}
+ \left(1-{\delta_{\mu 0}}\right){K_3}
\left({\partial_\mu}\phi - {A_\mu^I}\right)
\cr
\epsilon_{\mu\nu\lambda} \partial_{\nu}
{b_{\lambda}^I} \equiv {\partial_\mu}\phi - {A_\mu^I}
+ \left(1-{\delta_{\mu 0}}\right){K_3}
\left({\partial_\mu}\theta - {a_\mu} - {A_\mu^C}\right)
\cr
{\rho_{C,I}}\,{\rm Im}\left[{\Phi_{C,I}^*}
\left( {\partial_\mu} - i b_{\mu}^{C,I}  \right)
{\Phi_{C,I}}\right] \equiv \epsilon_{\mu\nu\lambda} \partial_{\nu}
\partial_{\lambda} \left\{\theta,\varphi\right\}
\end{eqnarray}
The right-hand-sides of the first two equations
are, repsectively, the conserved charge and pseudospin
currents of (\ref{L1}).
$\Phi_{C,I}$ create vortices in $\theta$ or $\phi$,
respectively. We have implicitly assumed that
vortices in $\theta_{1,2}$ (i.e. merons) are higher
in energy and can, therefore, be neglected.
As in the high-$T_c$ case \cite{Balents99},
this assumption leads to charge-pseudospin separation
and a particular form of topological order\cite{z2}.
We want to consider the more general situation
in which the lowest energy excitation is
a composite formed by $n$ vortices in $\theta$
and $m$ vortices in $\phi$,
${\Phi_{n,m}}\sim{\Phi_{C}^n}{\Phi_{I}^m}$, where
$n,m$ are integers. In this case,
we may write the effective action as:
\begin{eqnarray}
\tilde{{\cal L}}_D &=& 
\frac{1}{2} {\rho_{n,m}}\left| \left( \partial_{\mu} -
i n b_{\mu}^C - i m  b_{\mu}^I \right) {\Phi_{n,m}} \right|^2
\cr
& & +\,\frac{\kappa_0}{4} ( F^C_{\alpha\beta} )^2
 - \pi k i \epsilon_{\mu\nu\lambda}  b_{\mu}^C 
\partial_{\nu} b^C_{\lambda}
- i b_{\mu}^C \epsilon_{\mu\nu\lambda} \partial_{\nu} 
 A^C_{\lambda}
\cr
& & +\,\frac{\kappa_0}{4} ( F^I_{\alpha\beta} )^2
- i b_{\mu}^I \epsilon_{\mu\nu\lambda} \partial_{\nu} 
A^I_{\lambda} - \kappa_3 F^C_{0\alpha} F^I_{0\alpha}
\label{Lnm}
\end{eqnarray}

When there is no vortex condensate,
$\langle {\Phi_{n,m}} \rangle = 0$ for all $n,m$,
we may drop the vortex part of the action and
obtain the response functions from the remaining
(quadratic) terms in $b_{\mu}^C$, $b_{\mu}^I$:
\begin{eqnarray}
\sigma^{CC}_{xx} &=& 0 \hspace{2cm}
\sigma^{CC}_{xy} = \frac{1}{2 \pi k } \label{e12}\\
\sigma^{IC}_{xx} &=& 0  \hspace{2cm}
\sigma^{IC}_{xy} = \frac{\kappa_3}{ \kappa_0} 
\frac{1}{2  \pi k} \label{e34}\\
\sigma^{II}_{xx} &=& \frac{1}{\kappa_0} \frac{i}{\omega}  \hspace{1.5cm}
\sigma^{II}_{xy} = \left( \frac{\kappa_3}{ \kappa_0} \right)^2 
\frac{1}{2 \pi  k}
\label{e56}
\end{eqnarray}
Equations (\ref{e12}) tell us that we have an incompressible
quantum liquid with quantized Hall conductance. The first
of equations  (\ref{e56}) describes an interlayer superfluid
with a singularity at zero frequancy due to the Goldstone
mode associated with broken pseudospin symmetry.
It is remarkable that our simple analysis allows us to
calculate the weight of this mode as a function of charge imbalance 
between the layers
\begin{eqnarray}
I(\rho_1-\rho_2) = \frac{1}{\kappa_0} = I_0 \left[\, 1 
- { \left( \,\frac{ \rho_1 - \rho_2}{\rho_1 + \rho_2}\,
\right) }^2 \,\right] 
\label{Ir1-r2}
\end{eqnarray}
This is a simple prediction of our theory (which is,
in principle, easily verifiable experimentally)
of the spectral weight of the pseudospin Goldstone mode
for the usual $(k,k,k)$-type incompressible coherent phase
or the insulating compressible intra-layer coherent phase
in the unbalanced (${\rho_1}\neq{\rho_2}$) situation.

Various quantum disordered phases of the $(k,k,k)$ state
can also be described using (\ref{Lnm}). They result
when ${\Phi_{n,m}}$ condenses for some $n,m$.
When this occurs, the gauge field
$ n b_{\mu}^C + m  b_{\mu}^I$ aquires a gap.
After integrating it out, we obtain the response functions
of the corresponding quantum disordered phases
which we discuss below.

When $m=0$ and $n \neq 0$ we find that we destroy the quantum Hall 
effect without destroying interlayer coherence,
$ \sigma^{CC}_{xx}= \sigma^{CC}_{xy}= 0 $ and
$\sigma^{II}_{xx} = \frac{1}{\kappa_0} \frac{i}{\omega}$. 
So, relation (\ref{Ir1-r2}) will be satisfied even when
the state becomes compressible.
As we mentioned in our introductory comments, this
state may be relevant to the experiment of \cite{Spielman00}. 
Since the flux of $ b^C_{\lambda}$ is fixed by
$\epsilon_{0 \nu\lambda}  \partial_{\nu} b^C_{\lambda}
= {\rho_1} + {\rho_2}$, it must penetrate the
${\Phi_{n,0}}$ condensate. As a result, translational
symmetry must be broken, either spontaneously
by Wigner crystallization
or manifestly by disorder. In the perfectly
clean limit, the flux of  $b^C_{\lambda}$ enters the
${\Phi_{n,0}}$ condensate in an analogue of
the Abrikosov flux lattice. From (\ref{Lnm}), we
see that flux tubes in ${\Phi_{n,0}}$ carry flux $2\pi/n$
and, hence, charge $1/n$, equally distributed between
the two layers. Hence, this is a Wigner crystal of
charge $1/n$ quasiparticles which is coherent between the
two layers. In the strong disorder limit,
this can be viewed as a localized
phase of charge $1/n$ quasiparticles.

When $m \neq 0$, we find that interlayer coherence
is destroyed but the quantum Hall effect is not
\begin{eqnarray}
\sigma^{CC}_{xx} &=& 0 \hspace{2cm}
\sigma^{CC}_{xy} = \frac{1}{2 \pi k } \label{de12}\\
\sigma^{IC}_{xx} &=& 0  \hspace{2cm}
\sigma^{IC}_{xy} = \frac{n}{m} \,
\frac{1}{2  \pi k} \label{ne34}\\
\sigma^{II}_{xx} &=& 0  \hspace{2.0cm}
\sigma^{II}_{xy} = \left( \frac{n}{m} \right)^2 
\frac{1}{2 \pi  k}
\label{ne56}
\end{eqnarray}
At small frequencies, $\sigma^{II}_{xx}
\propto i \omega$ so this phase corresponds to an interlayer
insulator, rather than a superfluid.
We notice the following remarkable property
shared by interlayer coherent and incoherent
states: ${\sigma^{CC}_{xy}}{\sigma^{II}_{xy}} =
\left({\sigma^{IC}_{xy}}\right)^2$.
When interlayer coherence is destroyed, the
pseudospin Hall conductances are quantized.
Since a vortex corresponding to
${\Phi_{0,1}}, {\Phi_{0,2}},\ldots, {\Phi_{0,m-1}}$,
or no vortex at all
can be threaded along either of the
non-contractible loops of the torus,
there is an $m^2$-fold ground state
degeneracy in the pseudospin
sector which, when combined with the $k$-fold
degeneracy of the charge sector, gives a total
ground state degeneracy of $k m^2$ on the torus.
For general $n,m$, these states break translational
invariance. However, in the case $n=0$, $m \neq 0$,
$\epsilon_{0 \nu\lambda}  \partial_{\nu} b^I_{\lambda}
= {\rho_1} - {\rho_2}$, so translational symmetry will
only be broken if the layers are unbalanced,
in which case the charge difference between the
layers will be modulated.

We note that there is a straightforward
generalization akin to the hierarchy construction
which involves the introduction of an additional
Chern-Simons field, $\beta_\mu$, which attaches
$2l$ flux tubes to ${\Phi_{n,m}}$. Our earlier construction
was simply the $l=0$ case. The $l\neq 0$ modification
suggests a new set of
states with the same conductances as in
(\ref{de12}), (\ref{ne34}), but with $\sigma^{II}_{xy}$
(\ref{ne56}) replaced by $\sigma^{II}_{xy}
= ({1}/{2 \pi  k}) ({n^2 - 2lk})/{m^2}$.
It may be shown, using standard arguments
of CSLG theory \cite{Read90},
that these states have a ground state degeneracy of $k m^2$
on the torus; this is independent of $l$ and is
the same as in the $l=0$ case. When $m=0$, we obtain an
inter-layer coherent state with $\sigma^{CC}_{xy} =
2l/{2 \pi (2lk\pm 1)}$.

A physical interpretation for many of these
states may be given by considering a quantum Hall
state obtained by condensing composites which consist of $p$ electrons
in layer 1 and $q$ electrons in layer 2. We assume that
the composite is tightly-bound so that we can ignore its internal
structure. Let $\Psi_c$ be the operator
that creates the auxiliary boson defined by statistical transmutation
of such a composite. The CSLG theory for $\Psi_c$
takes the form
\begin{eqnarray}
{\cal L}_c &=& \Psi_c^{\dagger} ( \partial_0 - a_0 - p A_0^{(1)}
- q A_0^{(2)} ) \Psi_c
\cr & & + \,\frac{1}{2 m} \left| \left[ \vec{\partial} 
- i \vec{a} - i p \vec{A}^{(1)} - 
i q  \vec{A}^{(2)}\right] \Psi_c \right|^2 
\cr & & - \, \frac{i}{4 \pi k_c}\, \epsilon_{\mu\nu\lambda}
a_{\mu} \partial_{\nu} a_{\lambda}
\label{Lc}
\end{eqnarray}
If we take $k_c = k ( p+q )^2$, then we have a quantum
Hall state with $\sigma^{CC}_{xy} = 1/(2 \pi k )$.
To ensure that the composite has the correct statistics,
we must have $k$ odd.
For the particular case of quasiparticle pairs, this relation has
been discussed by  Halperin in the context of Laughlin's 
wavefunctions \cite{Halperin83}.
Proceeding to the dual representation of (\ref{Lc})
as earlier, we obtain
\begin{eqnarray}
{\cal L} &=& \frac{\kappa}{4} ( \tilde{F}_{\mu\nu} )^2
- i (p+q ) \, \tilde{b}_{\mu} \epsilon_{\mu\nu\lambda} 
\partial_{\nu}  A^C_{\lambda}
\cr & &- i (p-q)  \tilde{b}_{\mu} \epsilon_{\mu\nu\lambda} \partial_{\nu} 
A^I_{\lambda}
- \pi k (p+q)^2 i \epsilon_{\mu\nu\lambda}  \tilde{b}_{\mu} 
\partial_{\nu} \tilde{b}_{\lambda}
\label{Lcd}
\end{eqnarray}
where $\tilde{b}$ is the dual gauge field describing the currents of $\Psi_c$. 
Using (\ref{Lc}) or (\ref{Lcd}), we find
the response functions of the quantum Hall
state of composite objects 
$\sigma^{CC}_{xy} = \frac{1}{ 2 \pi k}$,
$\sigma^{IC}_{xy} = \frac{1}{ 2 \pi k} \, \frac{(p-q)}{ (p+q)  }$, and
$\sigma^{II}_{xy} = \frac{1}{ 2 \pi k} \, \frac{(p-q)^2}{  (p+q)^2 }$.
From the Chern-Simons theory (\ref{Lcd}), we deduce
a ground state degeneracy of $k (p+q)^2$ on the torus.
Hence, we find precisely the same conductance tensor 
and ground state degeneracy on the torus which we found earlier,
with $p+q = m$ and $p-q = n$.

Let us consider two special cases.
For $p=1$ and $q=0$, so that
quantum liquid is in one layer only, we have $m =1$,
and $n=1$, so it may be described by the condensation
of a composite formed by one vortex in $\theta$ and one
vortex in $\varphi$, i.e. a double vortex in $\theta_1$.
When $p=1$ and $q=1$  we have the paired state suggested in
\cite{Bonesteel96}. It is described by
$m = 2$ and $n=0$, i.e. by the condensation
of double vortices in $\varphi$. In this case, we
see from (\ref{Lnm}) that when
the layers are unbalanced, the charge difference
enters in a Wigner crystal of isospin $1/2$
quasiparticles (charge difference  between the layers
$e/2$).

To summarize, we have used a $U(1)$ particle-vortex duality
to extend a CSLG theory for $\nu=1/k$
(with $k$ an odd integer) for bilayer quantum Hall systems
to discuss the states  in which either Hall
quantization or interlayer coherence (``pseudospin suerfluidity'')
is individually destroyed (with the other still present) as
well as the more usual states in which both are present
(e.g. the $(1,1,1)$ state at $\nu=1$) or both are
absent. The compressible interlayer coherent states
are not translationally invariant and are therefore likely to be
disorder-driven localized states or Wigner crystal
(or CDW) states in each layer. Our most important new
conceptual results are the identification of
the theoretical possibility that there may be $\nu=1$ (or $1/k$)
bilayer pseudospin coherent states which are compressible
(unlike the usual $(1,1,1)$ state which is incompressible
and interlayer coherent) and the observation that the
experimental data presented in ref. \cite{Spielman00}
are {\it not} manifestly inconsistent with the exciting
prospect that such a pseudospin-coherent
compressible state (most likely a disorder-driven
Hall insulating phase) may actually be playing a
role in ref. \cite{Spielman00}.

It is a great pleasure to acknowledge discussions
with B. Halperin, J.P. Eisenstein, and N. Read.
CN and SDS acknowledge the Aspen Center for Physics.
ED is supported by
the Harvard Society of Fellows. CN is supported by the
National Science Foundation under Grant No. DMR-9983544
and the Alfred P. Sloan Foundation. SDS is supported
by the Office of Naval Research.

{

\end{document}
\vskip -0.6 cm}

\begin{thebibliography}{99}

\bibitem{Spielman00}
I.~B. Spielman, {\it et al.}, Phys. Rev. Lett. {\bf 84}, 5808 (2000).

\bibitem{Eisenstein97}
See, for example, the articles by J.~P. Eisenstein,
S.~M. Girvin and A.~H. MacDonald in {\it Perspectives
in Quantum Hall Effects}, editied by S. Das Sarma
and A. Pinczuk (Wiley, New York, 1997); and references therein.

\bibitem{Fertig89}
H.~A. Fertig, Phys. Rev. B {\bf 40}, 1087 (1989).

\bibitem{Wen92}
X.~G. Wen and A. Zee, Phys. Rev. Lett. {\bf 69}, 1811 (1992);
Phys. Rev. B {\bf 47}, 2265 (1993).



\bibitem{Ezawa92}
Z.~F. Ezawa and A. Iwazaki, Int. J. Mod. Phys.
B {\bf 6}, 3205 (1992).

\bibitem{He93}
S. He, {\it et al.}, Phys. Rev. B {\bf 47}, 4394 (1993);
Y.~W. Suen, {\it et al.}, Phys. Rev. Lett. {\bf 68}, 1379 (1992);
J.~P. Eisenstein,{\it et al.}, Phys. Rev. Lett. {\bf 68}, 1383 (1992).

\bibitem{Moon95}
K. Moon, {\it et al.}, Phys. Rev. B {\bf 51}, 5138 (1995);
K. Yang, {\it et al.}, Phys. Rev. B {\bf 54}, 11644 (1996).

\bibitem{Murphy95}
S.~Q. Murphy,{\it et al.}, Phys. Rev. B {\bf 52}, 14825 (1995).

\bibitem{Bonesteel96}
N.~E. Bonesteel,{\it et al.}, Phys. Rev. Lett. {\bf 77}, 3009 (1996).

\bibitem{DasSarma}
S. Das Sarma, {\it et al.}  Phys. Rev. Lett. {\bf 79}, 917 (1997); 
Phys. Rev. B {\bf 58}, 4672 (1998).

\bibitem{Stern00}
A. Stern,{\it et al.}, Phys. Rev. Lett. {\bf 84}, 139 (2000).



\bibitem{Schliemann00}
J. Schliemann, {\it et al.}, cond-mat/006309;
A. Stern, {\it et al.}, cond-mat/0006457;
L. Balents and L. Radzihovsky, cond-mat/006450;
M.~M. Fogler and F. Wilczek, cond-mat/007403.

\bibitem{Lee90}
D.-H. Lee and C.~L. Kane, Phys. Rev.  Lett. {\bf 64},
1313 (1990).

\bibitem{Kivelson92}
S. Kivelson, D.-H. Lee, and S.-C. Zhang,
Phys. Rev. B {\bf 46}, 2223 (1992).

\bibitem{Brey00}
L. Brey and H.~A. Fertig, cond-mat/0002218.

\bibitem{Balents99}
L. Balents, M.~P.~A. Fisher, and C. Nayak,
Phys. Rev. B {\bf 60}, 1654 (1999);
Phys. Rev. B {\bf 61}, 6307 (2000);
E. Demler, C. Nayak, H.-Y. Kee, Y.B. Kim,
and T. Senthil, in prep.


\bibitem{Zhang89}
S.~C. Zhang, {\it et al.}, Phys. Rev. Lett.
{\bf 62}, 82 (1989); N. Read, Phys. Rev. Lett.
{\bf 62}, 86 (1989).


\bibitem{dual} 
M.~P.~A. Fisher and D.-H. Lee,
Phys. Rev. B {\bf 39}, 2756 (1989), and references
therein.




\bibitem{Halperin83}
B.~I. Halperin, Helv. Phys. Acta {\bf 56}, 75 (1983).





\bibitem{z2}
This can be made more explicit by introducing operators
$\Phi_{1,2}$ which create vortices in $\theta_{1,2}$,
so that ${\Phi_{C}}={\Phi_{1}}{\Phi_{2}}$ while
${\Phi_{I}}={\Phi_{1}}{\Phi_{2}^*}$. Then,
our form of topological order
(charge-pseudospin separation)
is equivalent to the preservation of the $Z_2$ symmetry
${\Phi_{1,2}}\rightarrow -{\Phi_{1,2}}$.


\bibitem{Read90}
N. Read, Phys. Rev.  Lett. {\bf 65}, 1502 (1990);
X.~G. Wen and A. Zee, Phys. Rev. B {\bf 46}, 2290 (1992).


\end{thebibliography}
